\documentclass[journal]{IEEEtran}
\ifCLASSINFOpdf
\else
\fi
\pdfoutput=1
\usepackage{graphicx}
\usepackage{amsfonts}
\usepackage{amssymb}
\usepackage[cmex10]{amsmath}
\begin{document}
\title{Estimators of the correlation coefficient\\ in the bivariate exponential distribution}

\author{W.~J.~Szajnowski~\IEEEmembership{}
        {}
\thanks{W. J. Szajnowski, is with Centre for Vision, Speech and Signal Processing,  
  University of Surrey,  
Guildford, U.K.,
 e-mail: w.j.szajnowski@surrey.ac.uk}
\thanks{This work has been submitted to the IEEE for possible publication. Copyright may be
 transferred without notice, after which this version may no longer be accessible. 
 In such a case, however, a modified version of this work will be accessible.}
}

\markboth{ }%
{Shell \MakeLowercase{\textit{et al.}}: Bare Demo of IEEEtran.cls for Journals}
%



\maketitle

\begin{abstract}
\boldmath
 A finite-support constraint on the parameter space is used to derive a 
 lower bound on the error of an estimator of the 
  correlation coefficient in the bivariate exponential distribution.
 The bound is then exploited to examine optimality of three 
  estimators, each being a nonlinear function of moments of 
 exponential or Rayleigh observables.
   The estimator based on a measure of cosine similarity is shown to be  
  highly efficient for values of the correlation coefficient greater than 0.35; 
 for smaller values, however, it is the transformed Pearson correlation coefficient 
  that exhibits errors closer to the derived bound.
\end{abstract}


\IEEEpeerreviewmaketitle

\section{Introduction}
\IEEEPARstart{T}{he} bivariate exponential and Rayleigh probability distributions,  
 [1, pp. 401--475], [2], 
  play a prominent role in the development 
 of models of dependent nondeterministic phenomena in science and engineering. 
 Such models include 
 power of a random signal 
 received at multiple sensors exploiting time/space/frequency diversity, 
 weather radar returns observed 
 at co-polar and cross-polar channels, 
  weights of edges in random graphs being matched, time intervals between 
 significant events occuring in parts of a complex biological or man-made 
 system and many more.

 The statistical association between observables of interest can be characterized by 
 exploiting various measures of dependence, such as  mutual information, 
 copulas and parametric or nonparametric correlation coefficients 
  [1, pp. 105--177], [3]. In practice, 
  the correlation coefficient appears to be a preferred choice owing to
  its computational simplicity, and also     
 the fact that it can be functionally related to copulas and mutual information [3], [4]. 

The problem of estimating the correlation coefficient between non-negative observables  
 has been discussed in a number of publications [5]--[8]. However, since  
  the finite-support constraint on the parameter space has been ignored,  
 no conclusions regarding optimality of proposed estimators could be drawn. 
  Therefore, it is of interest 
 to establish a {\em {constrained}} lower bound on the estimator error and 
 examine estimators that could attain this bound.

\begin{figure*}
\center
\includegraphics[width=14.5cm]{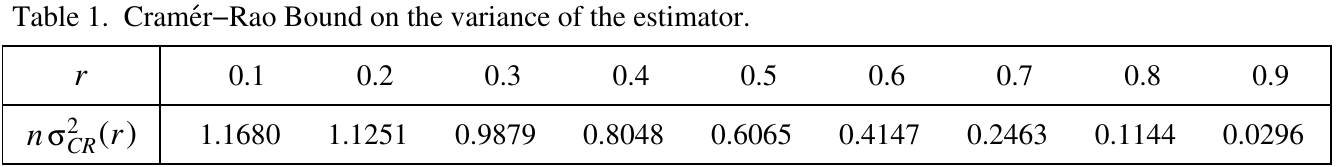}
\vspace{-0.4cm}
\end{figure*}

\vspace{0.5cm}

\section{Rayleigh and Exponential Distributions}
Consider two complex Gaussian random variables (rvs), $\mathbf{X} \triangleq X_I+jX_Q$ and 
$\mathbf{Y} \triangleq Y_I+jY_Q$, where $j^2=-1$. The four jointly Gaussian components, 
 $(X_I,X_Q,Y_I,Y_Q)$,
 have all zero means, 
 $\mathrm{E}\{X_I\}=\mathrm{E}\{X_Q\}=\mathrm{E}\{Y_I\}=\mathrm{E}\{Y_Q\}~=0$, 
where $\mathrm{E}\{\cdot\}$ denotes expectation, and their 
  covariance matrix is of the form [9]
\begin{equation}
 {\boldsymbol {\mathcal {C}}}_{XY} =
\begin{bmatrix}
\sigma^2_X & 0 & \sigma_X \sigma_Y \rho_c & \sigma_X \sigma_Y \rho_s\\
0 & \sigma^2_X & -\sigma_X \sigma_Y \rho_s & \sigma_X \sigma_Y \rho_c\\
\sigma_X \sigma_Y \rho_c & -\sigma_X \sigma_Y \rho_s & \sigma^2_Y & 0\\
\sigma_X \sigma_Y \rho_s & \sigma_X \sigma_Y \rho_c & 0 & \sigma^2_Y
\end{bmatrix}
\end{equation}
where $|\rho_c|\le 1$ and $|\rho_s|\le 1$ are correlation coefficients between respective rvs.

 In signal processing, the complex Gaussian rvs, $\mathbf X$ and $\mathbf Y$, 
 may be viewed as discrete-time samples of 
  two dependent complex Gaussian processes
 ${\mathbf X}(t)$ and ${\mathbf Y}(t)$. The rvs, $\mathbf X$ and $\mathbf Y$, may also  
 represent samples, taken at different time instants, say, $t$ and $t+\tau$, of
 a single stationary complex Gaussian process ${\mathbf X}(t)$; in such a case, 
 ${\mathbf Y}(t) = {\mathbf X}(t+\tau)$ and $\sigma^2_Y = \sigma^2_X$.

\subsection{Bivariate Rayleigh Distribution}
 Pairs of rvs, $(X_I,X_Q)$ and $(Y_I,Y_Q)$, can be used to construct 
 two Rayleigh rvs, $V$ and $Z$, as follows
\begin{equation}
V = \sqrt{X^2_I+X^2_Q} \quad \,\, {\rm and} \quad \,\, Z = \sqrt{Y^2_I+Y^2_Q}
\end{equation}
The rvs $V$ and $Z$ represent magnitudes of the corresponding 
 underlying complex Gaussian rvs $\mathbf X$ and $\mathbf Y$.

The joint probability density function (pdf) of $V$ and $Z$ is given by [2]
\begin{IEEEeqnarray}{rC}
p_{VZ}(v,z) = \frac{vz}{\sigma^2_X \sigma^2_Y(1-\rho^2)} 
\exp\!
 \left[- \frac {1}{2(1-\rho^2)} \left( \frac{v^2}{\sigma^2_X} + \frac{z^2}{\sigma^2_Y}\right)
\right] \nonumber \\
 \times\, I_0\! \left[\frac {\rho vz}{\sigma_X\sigma_Y(1-\rho^2)} \right]\!, 
\quad v, z \ge 0, \,\,\, \rho \ge 0  \,\,\,\quad
\IEEEeqnarraynumspace
\end{IEEEeqnarray}
 where $\rho^2 = \rho^2_c+\rho^2_s$, and $I_0(\cdot)$ denotes 
 a modified Bessel function of the first kind of order zero. If $\rho=0$, then  
 $p_{VZ}(v,z)\! =\! p_{V}(v)\mspace{2mu} p_{Z}(z)$, where $p_{V}(v)$ and $p_{Z}(z)$ 
 are marginal Rayleigh pdfs of 
  $V$ and $Z$, respectively. Therefore, in this case, zero correlation implies 
 statistical independence.

Population joint moments, $\mathrm {E}\{V^{\kappa}Z^{\nu}\}, \kappa\!+\!\nu=1,2$,
of rvs $V$ and $Z$ are given by [9]
\begin{IEEEeqnarray}{lll}
\mathrm {E}\{V\} = \sigma_X \sqrt{\pi/2}, \qquad  \,\, \,\,
 \mathrm{E}\{Z\} = \sigma_Y \sqrt{\pi/2}  
 \nonumber \\
\mathrm{E}\{V^2\} = 2 \sigma^2_X,  \qquad \qquad \,\,
\mathrm{E}\{Z^2\} = 2 \sigma^2_Y 
\nonumber \\
\mathrm{E}\{VZ\} = 
 \sigma_X \sigma_Y\!\! \left[ 2\mspace{1mu} {\mathsf E}(\rho) - (1-\rho^2){\mathsf K}(\rho)\right]
\end{IEEEeqnarray}
 where ${\mathsf K}(\cdot)$ and ${\mathsf E}(\cdot)$ are complete elliptic integrals 
of the first and second kind. In particular,
\begin{equation}
{\mathsf K}(0) = {\mathsf E}(0) = \pi/2, \quad {\mathsf E}(1) = 1 
\end{equation}
and when $\rho$ approaches one, ${\mathsf K}(\rho)$ tends to infinity.

\subsection{Bivariate Exponential Distribution}
The transformation
\begin{equation}
U = V^2 \quad  {\rm and} \quad  W = Z^2
\end{equation}
converts two Rayleigh rvs, $V$ and $Z$, into two exponential rvs, $U$ and $W$.
 The joint pdf of $U$ and $W$ can be expressed as [2]
\begin{IEEEeqnarray}{rC}
p_{UW}(u,w) = \frac{1}{4\sigma^2_X \sigma^2_Y(1-r)}
\exp\!
 \left[- \frac {1}{2(1-r)} \left( \frac{u}{\sigma^2_X} + \frac{w}{\sigma^2_Y}\right)
\right] \nonumber \\
 \times\, I_0\! \left[\frac {\sqrt{ruw}}{\sigma_X\sigma_Y(1-r)} \right]\!,
\quad u, w \ge 0, \,\,\, r \ge 0  \,\,\quad
\IEEEeqnarraynumspace
\end{IEEEeqnarray}
 where $r=\rho^2$. 
 The parameter $r$ is, in fact, the correlation coefficient between exponential rvs $U$
 and $W$ (see Section V).
 Also in this case, when $r=0$, rvs $U$ and $W$ are
 statistically independent.

Population joint moments, $\mathrm{E}\{U^{\kappa}W^{\nu}\}, \kappa\!+\!\nu=1,2$, 
 of rvs $U$ and $W$ are given by [9]
\begin{IEEEeqnarray}{lll}
 \mathrm{E}\{U\} = 2 \sigma^2_X,\quad \,\,\,\,\, \mathrm{E}\{W\} = 2 \sigma^2_Y
 \nonumber \\
\mathrm{E}\{U^2\} = 8 \sigma^4_X, \quad \,\, \mathrm{E}\{W^2\} = 8 \sigma^4_Y
 \nonumber \\
 \mathrm{E}\{UW\} = 4 (r+1) \sigma^2_X \sigma^2_Y.
\end{IEEEeqnarray}
 
\section{Problem Formulation}
Assume that observations on rvs $U$ and $W$ 
 are made in pairs,
$ (\mathbf{u}, \mathbf{w}) = \{(u_i,w_i): i = 1,2, \ldots, n\}$; alternatively, 
 observations, 
  $(\mathbf{v}, \mathbf{z}) =   \{(v_i,z_i): i = 1,2, \ldots, n\}$,
 may be made on Rayleigh rvs $V$ and $Z$.
Then, $n$ pairs of observations are used to determine sample joint moments,
\begin{equation}
m_{E\kappa\nu} \, \triangleq \, \frac {1}{n}\sum_{i=1}^n u^{\kappa}_i w^{\nu}_i 
    \quad {\rm or} \quad 
m_{R\kappa\nu} \, \triangleq \, \frac {1}{n}\sum_{i=1}^n v^{\kappa}_i z^{\nu}_i
\end{equation}
corresponding, respectively, to population moments (8) or (4).  

 This Letter addresses two associated problems:

 1. Given the pdf (7) and the constraint, $0\le r \le 1$, 
 derive a lower bound on the error of an estimator of the correlation coefficient $r$ 
 appearing in (7).

 2. Make use of sample moments (9) to construct estimators of $r$
 and examine their optimality with respect to the derived lower bound.

\section{Lower Bounds on Estimation Errors}
 In the case of a bivariate exponential distribution (7), allowed values of 
 the correlation coefficient $r$ are restricted to the $(0,1)$-interval.
 If a statistic employed as an estimator of $r$  
 assumes values from a different, finite or infinite, interval, then 
 the constraint, $0\le r \le 1$ must be 
 taken into account when establishing a lower bound on the estimator error.

\subsection{Cram\'er-Rao Bound (CRB)}
It is known [10] that under suitable regularity conditions, the variance of any {\em {unbiased}}  
 estimator can be bounded by the lower Cram\'er-Rao bound (CRB). Therefore, the CRB is 
 a useful measure when examining optimality of several competing estimators 
 of a parameter of interest.

Let a vector $\boldsymbol{\theta}$ of nonrandom parameters be defined by
\begin{equation}
\boldsymbol{\theta}\, \triangleq \, 
  (\theta_1,\theta_2,\theta_3)^T\, \equiv\, (r,\sigma^2_X,\sigma^2_Y)^T.
\end{equation}
Then (neglecting any constraints on the parameters), 
 the Fisher information matrix, $\boldsymbol{\mathcal I}(\boldsymbol{\theta})$, 
 is a $3\! \times\! 3$ positive semidefinite symmetric 
 matrix, comprising the elements
\begin{IEEEeqnarray}{lC}
\left[ \boldsymbol{\mathcal I}(\boldsymbol{\theta}) \right]_{k,\ell} \, \triangleq \, 
\mathrm{E} \left\{
\frac {\partial}{\partial\theta_k}\ln p_{UW}(u,w)\,\,\,
\frac {\partial}{\partial\theta_\ell}\ln p_{UW}(u,w)
 \right\}\!, \quad \nonumber \\ 
  \qquad \qquad \qquad \qquad \qquad \qquad \qquad \qquad \quad k,\ell = 1,2,3
\IEEEeqnarraynumspace
\end{IEEEeqnarray}
 Consequently, a lower bound on the variance of any unbiased estimator 
 $\hat{R}$ of $r$ can be determined from
\begin{equation}
\mathrm{var}\{\hat{R}\} \,\, \le \,\, 
 \frac {1}{n}
\left[ \boldsymbol{\mathcal I}^{-1}(\boldsymbol{\theta}) \right]_{1,1}
\, \equiv \, \, \sigma^2_{CR}(r),
\end{equation}
where $\boldsymbol{\mathcal I}^{-1}$ is the inverse of $\boldsymbol{\mathcal I}$.

Elements of the Fisher information matrix $\boldsymbol{\mathcal I}(\boldsymbol{\theta})$, 
 for selected values of $r$, are given in [11]. Values of the 
 Cram\'er-Rao bound, shown in Table 1,
 have been determined by selecting a first diagonal element of the inverse 
   $\boldsymbol{\mathcal I}^{-1}(\boldsymbol{\theta})$ of 
 $\boldsymbol{\mathcal I}(\boldsymbol{\theta})$.

\subsection{Mean-Square-Error (MSE) Bound}
 When the parameter space is restricted, the 
 Cram\'er-Rao approach appears to be inadequate [12]--[14]. Therefore,   
 to determine a lower bound on 
 the error of an estimator of parameter $r$ in (7), knowledge of the finite-support    
 constraint, $0\le r \le 1$, should be suitably combined with Fisher information contained in  
  available data.

Consider an unbiased estimator $\hat{R}$ 
 of $r$ and let 
 $p_{\hat{R}}(\hat{r};r)$ be a pdf of $\hat{R}$.
 Assume that the estimator $\hat{R}$ is so constructed that values of its realizations (estimates) 
 $\hat{r}$ cannot exceed one. 
 However, depending on a set of 
 processed data, $\{(u_i,w_i): i = 1,2, \ldots, n\}$ or $\{(v_i,z_i): i = 1,2, \ldots, n\}$, some 
 estimates $\hat{r}$  
  may assume negative, hence not allowed values.

 Therefore, when such an aberrant estimate $\hat{r}$ is observed, 
 its value must be set to zero, and 
 so modified estimate, $\hat{r}_m$, will assume the form
\begin{equation}
\hat{r}_m   \,\,=\,\, 
\left\{ \begin{array}{cl} {\!\!\hat{r},\quad}&{\hat{r} \geq 0}\\
                                      {\!\!0,\quad}&{\hat{r} < 0}. 
                   \end{array}
\right. 
\end{equation}
  
Consequently, the pdf of the modified estimator ${\hat{R}}_m$ will become a 
 {\em {censored}} distribution [15], 
\begin{equation}
p_{{\hat{R}}_m}({\hat{r}}_m;r) \, \,=\,\,
 \gamma\,\delta({\hat{r}}_m)
  \,+\, p_{{\hat{R}}}({\hat{r}}_m;r)\,\mathbf{1}({\hat{r}}_m)
\end{equation}
 comprising a discrete probability mass and a continuous part.
 In (14), $\delta(\cdot)$ is an impulse (Dirac delta) function, $\gamma$ is 
 the probability that ${\hat{R}} < 0$,
\begin{equation}
\gamma \,\, =\, \int_{-\infty}^0\! p_{\hat{R}}(\hat{r};r)\,d{\hat{r}}
\end{equation}
 and $\mathbf{1}(\cdot)$ 
 denotes the Heaviside step function,
\begin{equation}
\mathbf{1}(\omega) \,\, \triangleq \,\,
\left\{ \begin{array}{cl} {\!\!1,\quad}&{\omega \geq 0}\\
                                      {\!\!0,\quad}&{\omega < 0}.
                   \end{array}
\right. 
\end{equation}

Fig. 1 illustrates the effect of transforming the pdf of an estimator $\hat{R}$ into its  
 censored version, $p_{{\hat{R}}_m}({\hat{r}}_m;r)$, when the value of the correlation
 coefficient $r$ being estimated 
 decreases from $r_{\beta}$ to $r_{\alpha}$.

\begin{figure}[]
\centering
\includegraphics[width=7.cm]{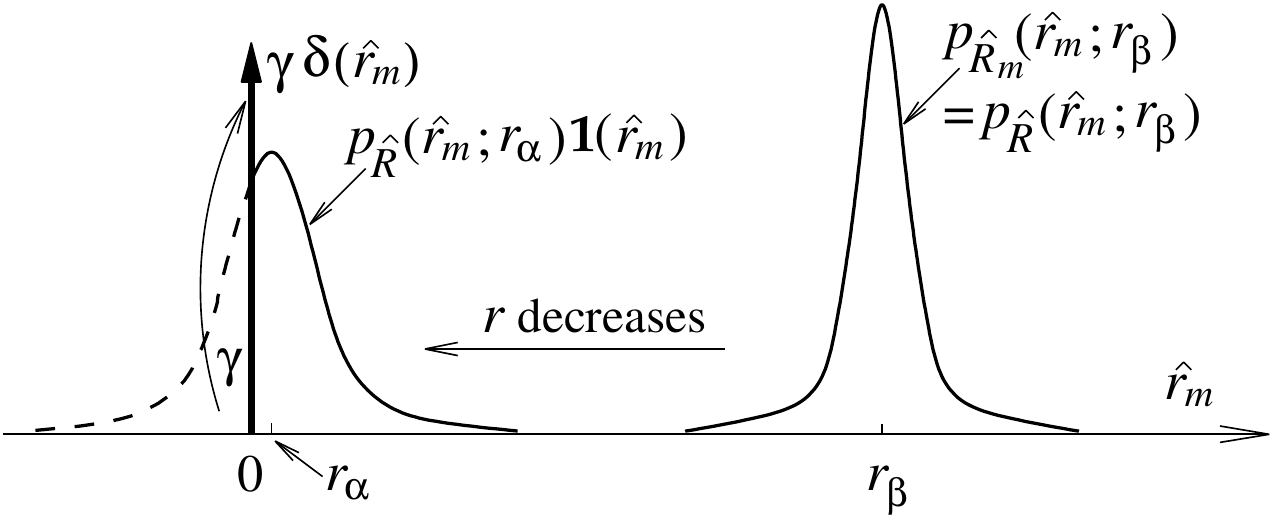}
\vspace{-0.2cm}
\caption{The effect of censoring when $r$ decreases from $r_{\beta}$ to $r_{\alpha}$.}
\end{figure}

 The mean-square error (MSE), $\varepsilon^2_{MS}(r)$, of the modified estimator
 $\hat{R}_m$ can be expressed as
\begin{equation*}
 \varepsilon^2_{MS}(r)  
 \,\triangleq \, \mathrm{E}\{(\hat{R}_m-r)^2\}  \,= \,
   \mathrm{var}\{\hat{R}_m\} + 
   {\underbrace{(\mathrm{E}\{\hat{R}_m\} - r)}_{\rm{bias\,\,squared}}}\mspace{-1mu}{}^2.
\end{equation*}

 In order to determine the lower MSE bound, 
 assume that $\hat{R}$ is a maximum-likelihood (ML) estimator. 
 Since ML estimators are known to be 
 asymptotically unbiased, efficient and 
 Gaussian [10], 
 let $\hat{R} \sim \mathcal{N}(r,\sigma^2_{CR})$.
 The MS error of a modified estimator $\hat{R}_m$ can be evaluated by exploiting 
 moments of a censored Gaussian distribution [15].

 Let $\phi(\lambda)=(1/\sqrt{2\pi}) \exp(-\lambda^2/2)$ be the pdf 
 of a standard Gaussian rv $\Lambda$, 
 and $F(\mu)=\Pr\{\Lambda \le \mu\}$ its cumulative distribution function.
 Then, the  MSE of the estimator $\hat{R}_m$ can be expressed as follows
\begin{IEEEeqnarray}{lC}
\varepsilon^2_{MS}(r) \,\,=\,\,
\underbrace{\sigma^2_{CR}F(\mu)\left[ (1-d) + F(-\mu)(\mu+h)^2\right]}_{\rm{variance}}
\,+ \nonumber \\
  \qquad \qquad \quad \,+  \, \underbrace{[F(\mu)\mspace{1mu}(r + h\mspace{1mu} \sigma_{CR}) - r]}
_{\rm{bias\,\,squared}}\!{}^2  
\end{IEEEeqnarray}
 where $\sigma^2_{CR} \equiv \sigma^2_{CR}(r),\,\,\mu = r/\sigma_{CR},\,\,h = \phi(\mu)/F(\mu)$ and  
  $d = h(h + \mu)$.

The constrained error bound (17) differs from the CR bound (12), 
  when $r$ is less than approximately $3\,\sigma_{CR}(r)\,$. 
In the region, $0 \le r < 3\,\sigma_{CR}(r)$, 
 the estimator $\hat{R}_m$ becomes {\em {biased}}, and its MS error,
\begin{equation}
   \sigma^2_{CR}(0)/2 \,\, \le \,\, \varepsilon^2_{MS}(r) \,\, < \,\, \sigma^2_{CR}(r)
\end{equation}
  remains {\em {below}} the CR bound. The bound reduction has resulted from 
 incorporating knowledge of the constraint.

\section{Estimators of the Correlation Coefficient}
Consider the population Pearson product-moment correlation coefficient defined by
\begin{equation}
r_P(U,W) \, \triangleq \, \frac {\mathrm{E}\{UW\}- \mathrm{E}\{U\}\mathrm{E}\{W\}}
 { \sqrt{ \mathrm{var}\{U\} \, \mathrm{var}\{W\} }}.
\end{equation}
 By inserting moments (8) into (19), it can be verified that 
$r_P(U,W) = r$.
  Therefore, the
 {\em sample} Pearson
 correlation coefficient, i.e. the statistic 
\begin{equation}
 \mathsf{s}(\mathbf{u},\mathbf{w})  \, = \,
\frac {m_{E11} - m_{E10}\,m_{E01}} {\sqrt{(m_{E20}-m^2_{E10})(m_{E02}-m^2_{E01}) }}
\end{equation}
 can be used to construct 
 a {\em {censored}} estimate $\hat{r}_1$ of $r$ as follows
\begin{equation}
\hat{r}_1   \,\,=\,\,
\left\{ \begin{array}{cl} 
{\!\!\mathsf{s}(\mathbf{u},\mathbf{w}),\quad}&{\mathsf{s}(\mathbf{u},\mathbf{w}) \geq 0}\\
                                      {\!\!0,\quad}&{\text{otherwise.}}
                   \end{array}
\right.
\end{equation}
 When the number $n$ of observations tends to infinity,
 sample moments converge to population moments, and
 the sample correlation
 coefficient $\mathsf{s}(\mathbf{u},\mathbf{w})$ will approach $r$.

 The use of sample correlation coefficient to estimate  
   a population correlation coefficient is a standard practice. However, such an approach 
 may not necessarily lead to 
 an efficient estimator (an estimator whose variance attains the Cram\'er-Rao bound), 
 especially in small or moderate sample sizes.

\subsection{Estimator Based on Correlation of Rayleigh Variables}
 Consider now the bivariate Rayleigh distribution (3) and the population Pearson 
 correlation coefficient $r_P(V,Z)$, 
 given by a formula analogous to (19). 
 The correlation coefficient $r_P(V,Z)$ can be expressed in terms of moments (4) 
 as follows
\begin{equation}
r_P(V,Z) \, = \, \frac 
{2\left[ 2\mspace{1mu} {\mathsf E}(\!\sqrt{r}\mspace{1mu}) - 
 (1-r){\mathsf K}(\!\sqrt{r}
  \mspace{1mu})\right]- \pi}
    {4 - \pi}.
\end{equation}
 In this case, $r_P(V,Z)=r$, only when $r=0$ or $r=1$; otherwise, $r_P(V,Z)$ is a
 nonlinear function of $r$. 

When $n \rightarrow \infty$, the sample correlation 
 coefficient $\mathsf{s}(\mathbf{v},\mathbf{z})$ will approach (22). 
By employing the nonlinear transformation
\begin{equation}
\xi(\mathbf{v},\mathbf{z}) \, = \, \mathsf{s}(\mathbf{v},\mathbf{z})
\left\{1 + g\mspace{1mu}[1-\mathsf{s}(\mathbf{v},\mathbf{z})] \right \}\!, 
\,\,\, g=49/500
\end{equation}
a censored estimate $\hat{r}_2$
 of $r$ is obtained as
\begin{equation}
\hat{r}_2   \,\,=\,\,
\left\{ \begin{array}{cl}
{\!\!\xi(\mathbf{v},\mathbf{z}),\quad}&{\xi(\mathbf{v},\mathbf{z}) \geq 0}\\
                                      {\!\!0,\quad}&{\text {otherwise.}}
                   \end{array}
\right.
\end{equation}

\subsection{Approximate Maximum-Likelihood Estimator}
 It has been shown [16] that in a case of highly correlated Rayleigh rvs, 
 and when $\sigma_X=\sigma_Y$,   
 an approximate ML estimator of 
 the correlation coefficient $\rho$ is of the form, 
 $[\mspace{1mu}2m_{R11}/(m_{R20}+m_{R02})]^2$. The constraint, $\sigma_X=\sigma_Y$, can be removed 
 by employing the geometric mean rather than the
 arithmetic mean. Consequently, 
  the following statistic of {\em {cosine-similarity}}-squared is obtained
\begin{equation}
 \mathsf{c}^2(\mathbf{v},\mathbf{z}) \: \triangleq \: 
\frac {m^2_{R11}} {m_{R20}\,m_{R02}}.
\end{equation}
 The statistic (25) asymptotically converges to
\begin{equation}
 \lim_{n\rightarrow \infty} \mathsf{c}^2(\mathbf{v},\mathbf{z}) \:=\: \left[
 {\mathsf E}(\!\sqrt{r}\mspace{1mu}) -
 (1-r){\mathsf K}(\!\sqrt{r}
  \mspace{1mu})/2 \right]^2.
\end{equation}
 For $r=0$ and $r=1$, 
 the respective limits are $\pi^2/16$ and $1$.

When the nonlinear transformation
\begin{IEEEeqnarray}{lC}
\eta(\mathbf{v},\mathbf{z}) \, = \, \frac {\mathsf{c}^2(\mathbf{v},\mathbf{z})-a}{1-a}\!\!\!\!\! 
 &\!\!\!\!\!\!\left\{1 + b[1-\mathsf{c}^2(\mathbf{v},\mathbf{z})] \right \}\!,\\
 &\qquad \,\,\,a= \pi^2/16, \,\,\, b= 7/12 \nonumber
\end{IEEEeqnarray}
is applied,  
a censored estimate $\hat{r}_3$
 of $r$ assumes the form
\begin{equation}
\hat{r}_3   \,\,=\,\,
\left\{ \begin{array}{cl}
{\!\!\eta(\mathbf{v},\mathbf{z}),\quad}&{\eta(\mathbf{v},\mathbf{z}) \geq 0}\\
                                      {\!\!0,\quad}&{\text{otherwise.}}
                   \end{array}
\right.
\end{equation}
Owing to its origin, the estimator $\hat{R}_3$  
 is expected to be asymptotically efficient, at least 
 for larger values of $r$.  

\subsection{Performance of the Estimators}
Computer simulations were employed to examine the performance of the three estimators, 
$\hat{R}_1,\,\hat{R}_2$ and $\hat{R}_3$, of the correlation coefficient $r$. Three sample 
 sizes, $n=10,\,n=50$ and $n=200$, were chosen, somewhat arbitrarily, to represent the   
 cases of small, moderate, and large sample sizes. Values of 
 the correlation coefficient $r$ to be estimated varied from 
 $r=0$ to $r=0.98$, in steps of $0.02$. For each combination of $n$ and $r$, $10^6$ 
Monte Carlo experiment replications were carried out to determine 
 the MS error, $\varepsilon^2$, for each of the three estimators.

\begin{figure}[]
\centering
\includegraphics[width=9cm]{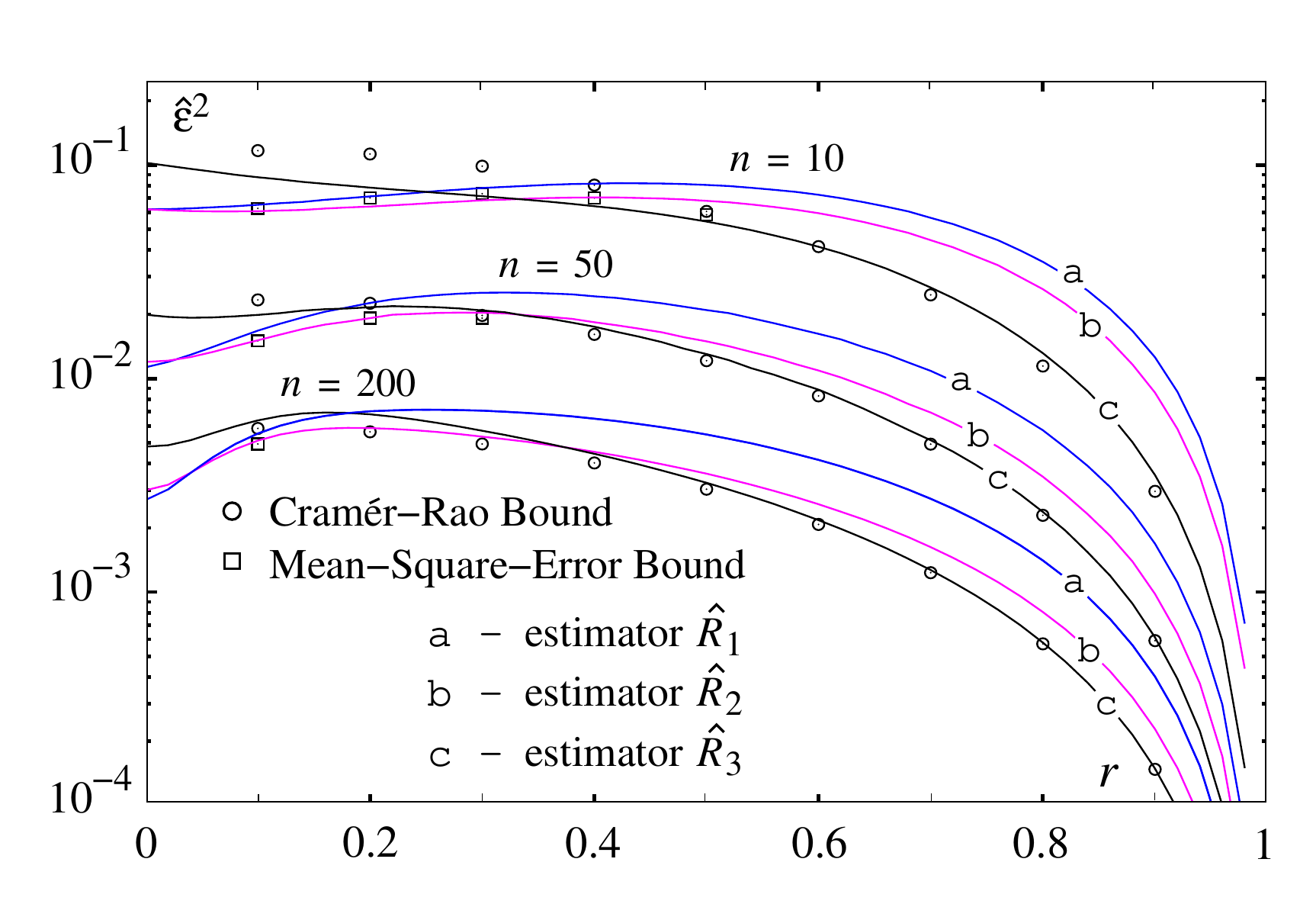}
\vspace{-0.5cm}
\caption{Estimated mean-square error, $\hat{\varepsilon}^2$, of the three estimators.}
\end{figure}

 Results of the 
 study are shown in Fig. 2 along with the MSE bound (17)
 and the Cram\'er-Rao bound (12); 
 values of the MSE bound are only shown when they differ from those of the CR bound.
 
 The results can be summarized as follows:

 1. The derived MSE lower bound is superior to the standard CRB when predicting errors of 
 estimators of the correlation coefficient $r$; the MSE lower bound  
  is more precise when the sample size is moderate or large.

 2. When $r$ is greater than $r^{*} \approx 0.35$, the estimator $\hat{R}_3$ is better than 
 the other two estimators, and its estimated MS error, $\hat{\varepsilon}^2$, 
 differs only slightly from the derived lower bound.

 3. In the region, $ r < r^{*}$,  
  the estimator $\hat{R}_2$ is superior to the estimator $\hat{R}_3$.

 4. When $r \approx 0$, the estimator $\hat{R}_1$ 
 exhibits the smallest MS error; this observation supports the conclusion in [8] 
 that the sample correlation coefficient (20) is an asymptotically most powerful test of the hypothesis 
 $r = 0$ against the alternative $r > 0$.

 5. When $r < 0.1$, the MS error of the estimator $\hat{R}_3$ markedly 
 exceeds those of the other two 
 estimators; this effect can partly be attributed to the approximate nature of the nonlinearity (27).

\section{Conclusion}
 The non-negativity constraint has been incorporated into the standard CR bounding technique  
 by utilizing moments of a censored Gaussian distribution. The resulting 
 MSE bound establishes a lower 
 bound on the MS error of any estimator of 
 the correlation coefficient of exponentially distributed variables.

 The simulation study has shown that MS errors associated with two of the examined 
 estimators are close to the derived lower bound in two subintervals that jointly cover the 
 entire (0,1)-interval. Each of the two estimators is a 
 nonlinear function of a measure of either cosine similarity or 
 {\em {centred}} cosine similarity  
 (i.e. the sample correlation coefficient) between Rayleigh variables.

\end{document}